\documentclass{mem}
\usepackage{natbib}\usepackage{txfonts}\usepackage{balance}
\usepackage{graphicx}
\usepackage[a4paper]{hyperref}
\idline{75}{282}
\begin{document}
\def\teff{$T\rm_{eff }$}
\def\kms{$\mathrm {km s}^{-1}$}
\def\gsimeq
{\hbox{\raise0.5ex\hbox{$>\lower1.06ex\hbox{$\kern-1.07em{\sim}$}$}}}
\def\lsimeq
{\hbox{\raise0.5ex\hbox{$<\lower1.06ex\hbox{$\kern-1.07em{\sim}$}$}}}

\title{Outflows and winds in AGNs: a case for Simbol-X}

   \subtitle{}

\author{M. \,Cappi\inst{1} 
\and F. \,Tombesi\inst{1,2}
\and M. \,Giustini\inst{1,2}
}

  \offprints{cappi@iasfbo.inaf.it}
 
\institute{
INAF-IASF, Bologna, 
Via Gobetti 101, 40129, Bologna, Italy
\and
Dipartimento di Astronomia, Universit\`a di Bologna, 
Via Ranzani 1, 40127, Bologna, Italy
\email{cappi@iasfbo.inaf.it}
}

\authorrunning{Cappi}

\titlerunning{Outflows and winds in AGNs}

\abstract{Chandra and XMM-Newton X-ray observations are accumulating evidence for massive, 
high velocity outflows in Seyfert galaxies and quasars, most likely 
originating very close to the central supermassive black hole.
These  results are offering new potential to probe the launching 
regions of relativistic jets/outflows, and to quantify their feedback 
impact on the host galaxy and/or cluster of galaxies. 
The most important signature of these phenomena is the detection of blueshifted absorption 
lines of highly ionized iron at energies usually greater than $\sim$ 7 keV. The lack of 
sensitivity of present day X-ray observatories at 
these energies gives rise to bias against the detection of more ``extreme'' outflows, with 
highest velocity and ionization, which would be blueshifted at energies $>$ 10 keV. 
Thus, simulations with Simbol-X were carried out to test its capability 
of detecting absorption lines/edges between 5-20 keV, in order to probe 
the dynamics (i.e. measurement of velocity variations) of the absorbing gas, as well as the 
highest (up to relativistic speeds) velocity and ionization components.
We found that the unprecedented sensitivity of Simbol-X between 5-30 keV is a
great opportunity to obtain important improvements in this research field. 
}
\maketitle{}

\section{Introduction}

Manifestations of fast winds/outflows/ejecta in AGNs are seen at all wavelengths: from 
the relativistic jets in radio-loud AGNs (Bridle \& Perley, 1984), to the optical/UV broad absorption lines 
detected in BAL QSOs (Weymann et al. 1991; Reichards et al. 2003), and the warm absorbers that are almost 
ubiquitous in X-ray bright AGNs (Crenshaw, Kraemer and George 2003).
Detailed studies from soft X-ray grating observations indicate multiple ionization and 
kinetic components with velocities up to few 1000 km s$^{-1}$ (Blustin et al. 2005).
Understanding how winds forms and what are their physical characteristics is of 
fundamental importance to estimate their energetic impact onto ISM and IGM.
Currently there are still order of magnitude uncertainties on the mass outflow rates involved
(see the review by Elvis (2006) and references therein).

New results from Chandra and XMM-Newton (see Cappi 2006 for a review and references therein) 
show blue-shifted absorption lines which indicate very high velocities, even larger than the 
highest velocities ($\sim$0.2 c) ever observed in the optical/UV and implying larger 
than expected outflow rates (King \& Pounds 2003).
The variability often associated to these absorption lines locates the absorbing gas 
very near the nuclear BH, potentially at the base of a wind from an accretion disk or 
at the base of a jet.

\section{On the possibility of foreground contamination}

It has been claimed that some of the most extreme outflows, in particular where (as is often the case) 
only one line is detected, may have been misidentified with ionized absorption produced locally in 
our Galaxy (Mc Kernan et al., 2004, 2005). To test this hypothesis, we have searched for a spatial 
correlaton between the sources with high-velocity outflows detected (taken from Cappi 2006) 
and any Galactic structures known to harbour metals and/or hot ionized gas such as Supenovae Remnants, 
the Local Bubble, the North Polar Spur, or similar.
Figure 1 shows the AGN locations overlaid to the (a) 3/4 keV emission map obtained from the 
RASS (Snowden et al. 1997), which traces (at least part of) the hot diffuse Galactic emission, 
and to (b) the major known Galactic structures. These figures show that no clear correlation is found.

\begin{figure}[]
\includegraphics[width=6truecm,height=4truecm,angle=0]{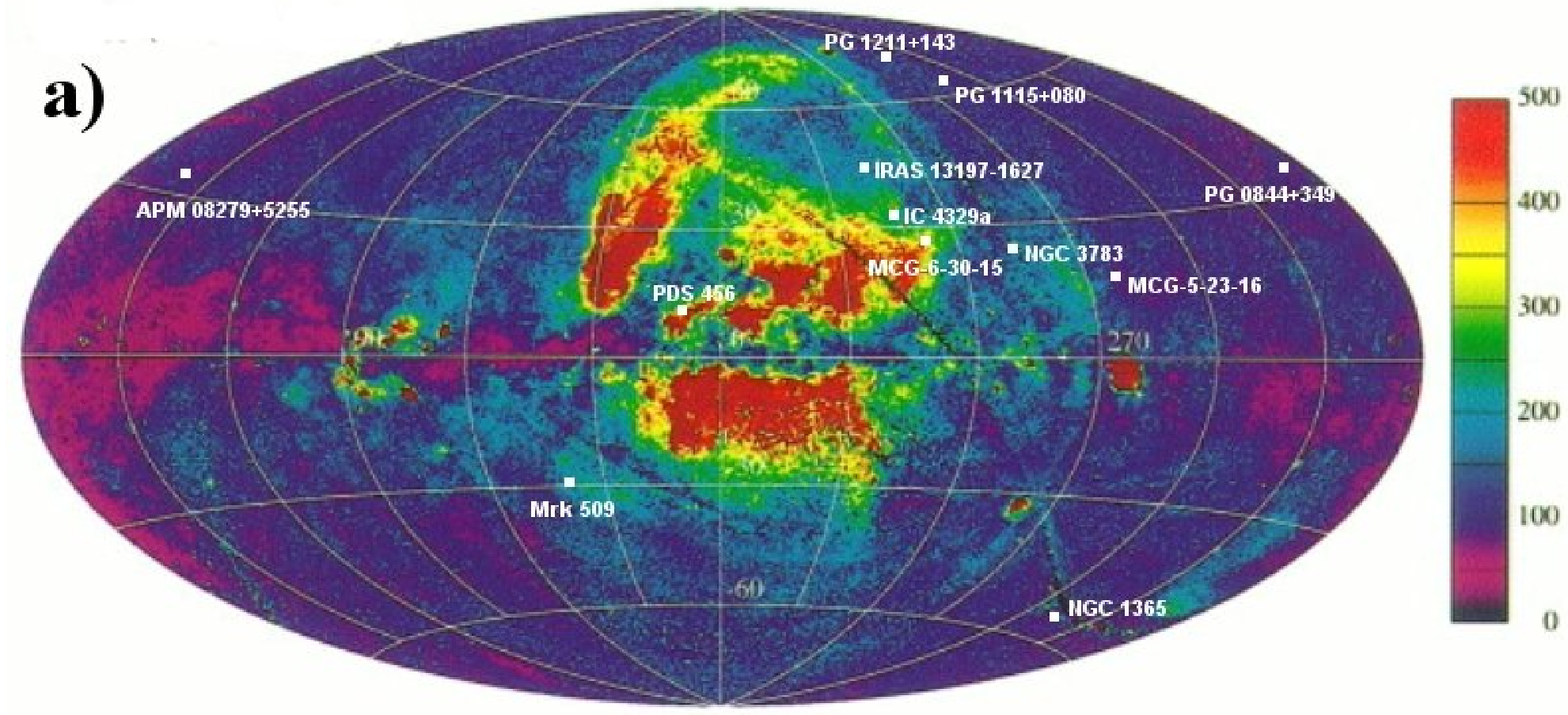}
\\
\includegraphics[width=6truecm,height=4truecm,angle=0]{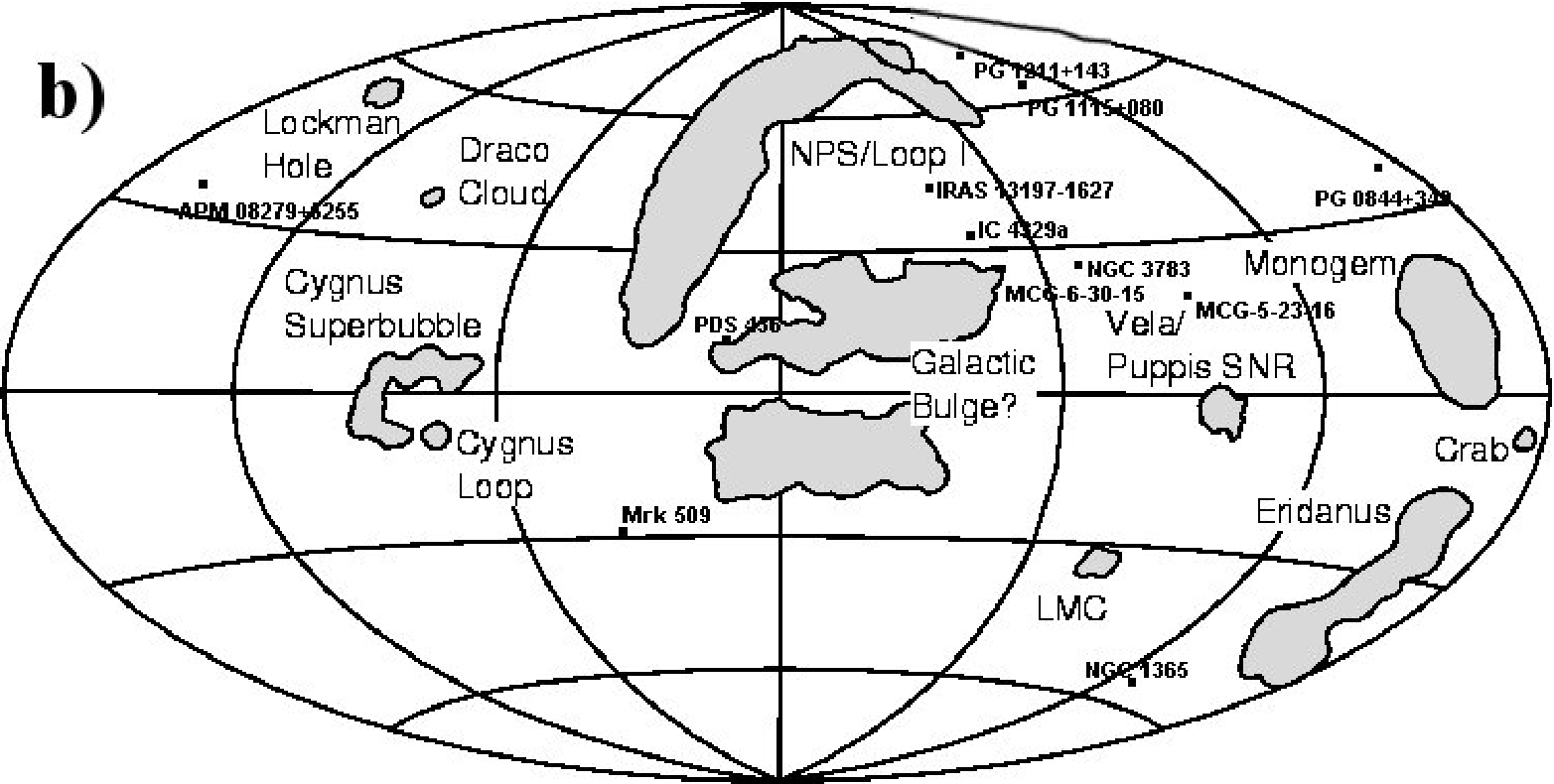}
\caption{\footnotesize  Location of AGNs with detected relativistic outflows (from Cappi 2006) superimposed on:
(a) the RASS map at 0.75 keV (Snowden et al. 1997). Color scale is in units of 10$^{-6}$ counts/s/arcmin$^2$; 
(b) the major known Galactic structures. 
}
\label{eta}
\end{figure}

Moreover, following McKernan et al. (2004, 2005), we have plotted (Figure 2) the energy blue-shift of the 
absorption lines as a function of the source cosmological redshift. If due to local contamination, the two parameters 
should follow a one-to-one linear correlation which is not, however, seen in the data. 
Alternatively, contamination may be due to a very tenuous and pervasive medium such a Warm-Hot Intergalactic 
Medium (WHIM). It could be too tenuous to emit a detectable amount of bremsstrahlung radiation, but could be of 
high enough colun density (once integrating over a much longer than Galactic path lenght) to produce 
significant X-ray absorption. However, under this hypothesis, X-ray absorption lines should be detected in the 
spectrum of most extragalactic objects, including blazars and/or clusters of galaxies, etc., which is not the case.
Moreover, in the few cases in which variability is detected on timescales of a few ks, the local diffuse origin  
would be clearly unlikely. We conclude that we do not find any evidence of Galactic or foreground 
contamination, strenghtening the connection between blueshifted absorption lines and outflows intrinsic 
to the AGNs.

\begin{figure}[!]
\includegraphics[width=6truecm,height=4truecm,angle=0]{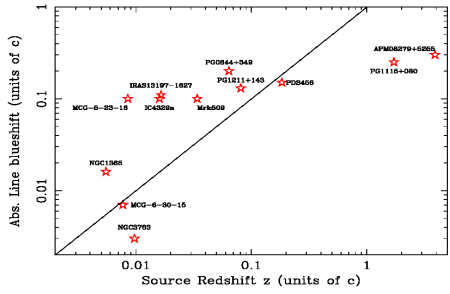}
\caption{\footnotesize
Blueshift of AGNs absorption lines versus source cosmological redshift. Errors on the absorption blueshift measurements 
are typically at a few $\%$ level, i.e. smaller than the symbols plotted in the figure. The straight line 
indicates the local standard of rest. 
}
\label{eta}
\end{figure}

It should be stressed that the FeK absorption lines at the highest energies 
have, up to now, suffered a number of biases against their detection. First, there is an 
``observational bias'' against the highest-velocity (bluest) components 
in that orbiting telescopes are of limited sensitivity at energies greater than 7 keV. 
Then, the sporadic nature of these features has also likely generated a ``detection bias''. 
Finally, detecting gas with the highest velocities, i.e. likely highest ionization states, 
restricts our studies to a few ionization levels of Fe, rather than a ``pletoria'' of low-Z elements.
All this, combined to the fact that high velocity, ionized and variable absorbing gas 
is naturally expected in models involving winds/ejecta/outflows (such as those 
predicted in MHD simulations, i.e. Proga 2005, Kato et al. 2004), suggests that the 
parameter space for new discoveries in this field is large, i.e. we may have seen up to 
now only the tip of the iceberg!
Two possible future directions for Simbol-X are thus addressed below.

\section{Simbol-X capabilities}

One among the most debated questions of modern astrophysics is to 
quantify the energy feedback of AGNs into the ISM and IGM. To do so, one 
important piece of information would be to have a detailed physical description 
(in terms of ionization state, geometry, kinematics, dynamics, mass outflow rate, etc.) 
of the outflows/winds that are known to be present in most AGNs, including QSOs 
(see review by Elvis 2006).
X-ray sensitive observations are mandatory to achieve this goal because they can, at least 
in principle, probe directly (through absorption spectroscopy) the highest ionization and velocity outflowing 
gas which may carry most of the kinetic energy. 

\begin{figure}[!]
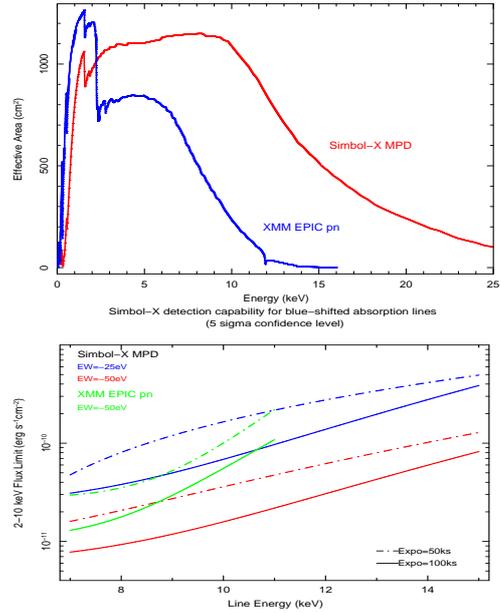

\includegraphics[width=4truecm,height=6.5truecm,angle=-90]{cappi_fig3a.ps}
\includegraphics[width=4truecm,height=7truecm,angle=-90]{cappi_fig3b.ps}
\caption{\footnotesize Comparison between XMM-Newton EPIC-pn and Simbol-X low energy detector (MPD). 
({\it Top}): Comparison of the effective areas: the MPD will have an effective area equivalent 
($\simeq$ 30\% better) than the pn 
at energies between $\sim$ 4-6 keV and from 2 to 5 times higher at energies greater than 7 keV; ({\it down}): 
Comparison of the sensitivity 
to absorption lines: the 2--10 keV flux limit is plotted against energy for the detection of an absorption 
line at 5 $\sigma$ confidence level. Equivalent widths are $-$25 and $-$50 eV, exposure times of 50 and 100 ks and background 
level at nominal value.} 
\label{eta}
\end{figure}

As shown below, we suggest to make use of the unprecedented throughput of Simbol-X (Ferrando et al. 2004) 
between 2-20 keV energy band (see Top panel in Figure 3 for a comparison with the pn onboard XMM-Newton) to obtain 
a great step forward in this direction.
In particular, two main topics could be addressed in detail with Simbol-X: i) the study of the dynamics of the 
highly ionized gas, through detailed study of the time evolution of the spectral absorption structures, and 
ii) the study of the highest velocity components of these outflows, possibly up to 
relativistic speeds.

Simulations of Simbol-X spectra have been performed using the latest 
response matrices and background files available\footnote{http://www.iasfbo.inaf.it/simbolx/faqs.php}. 
Lower panel in Figure 3, taken from Tombesi et al. 2007, shows the results obtained from a number of 
Simbol-X simulations aimed at estimating the Simbol-X sensitivity at detecting absorption lines, 
to be compared to that of XMM-Newton. It is found that Simbol-X will be 2 to 5 times more sensitive than 
the pn onboard XMM-Newton, and will also allow high detection capability up to 15 keV. 
This will allow detection of absorption lines down to EW of few tens eV for several tens of known X-ray emitting 
AGNs. It should be noted that these sensitivities, for sources brighter than few $\times$ 10$^{-11}$ erg/cm2/s 
only weakly depend on the detailed background level that is assumed.

Based on our current knowledge of AGN winds build upon Chandra and XMM-Newton experience, 
we have then simulated two ``textbook'' examples of AGNs with strong warm absorbers: the Seyfert 1.8 NGC1365 
(Risaliti et al. 2005), and the QSO PDS456 (Reeves et al. 2003).
We have compared in Fig. 4 the real XMM-Newton observation of NGC1365 
with a Simbol-X simulation of the same source, with the same exposure time. 
Clearly Simbol-X will be at least 2 to 5 times better than XMM-Newton in detecting absorption
lines up to 10 keV, and may also allow to detect features (like absorption lines 
and/or edges) up to 15 keV.

\begin{figure}[!]
\includegraphics[width=5truecm,height=6truecm,angle=-90]{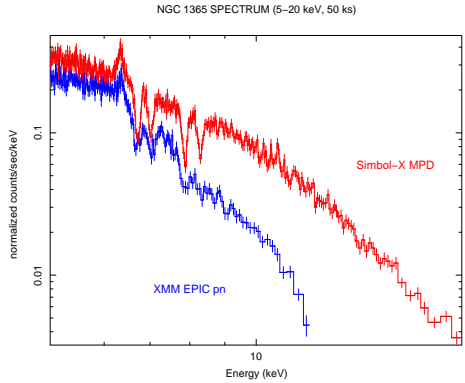}
\caption{\footnotesize The 5-10 keV spectrum of the Seyfert 1.8 NGC1365 
(F$_{(2-10 keV)}$$\sim$10$^{-11}$ ergcm$^{-2}$s$^{-1}$) clearly
showing four absorption lines due to FeXXV and FeXXVI K${\alpha}$ and 
K$\beta$, outflowing with velocities varying between $\sim$ 1000 and 
$\sim$ 5000 km/s among the observations performed by XMM-Newton (Risaliti et al. 2005).
Exposure time is 50 ks.
The spectrum is compared to a simulated Simbol-X spectrum with same exposure. 
}
\label{eta}
\end{figure}

Figure 5 shows the Simbol-X simulation of the time-energy spectral map 
obtained for a source like NGC1365 between 5-14 keV, during a 200 ks observation. Time intervals are binned at 2 ks
and energy bins at 100 eV.
The model consists of the four absorption lines detected in NGC1365 and described in Figure 4, 
plus a constant narrow emission FeK line. The ionized absorber is assumed to accelerate 
from v=0.02 c (as measured by XMM-Newton) up to 0.2 c during the first 100 ks, and 
decelerate from 0.2 c down to 0.02 c during the last 100 ks.
Clearly, Simbol-X will open up the possibility to follow the 
outflow time evolution (i.e. acceleration/deceleration) on 
time-scales (few ks) comparable to the dynamical timescale at a few Schwarzshild
radii from the black hole (Figure 5).

\begin{figure}[!]
\includegraphics[width=6truecm,height=5truecm,angle=0]{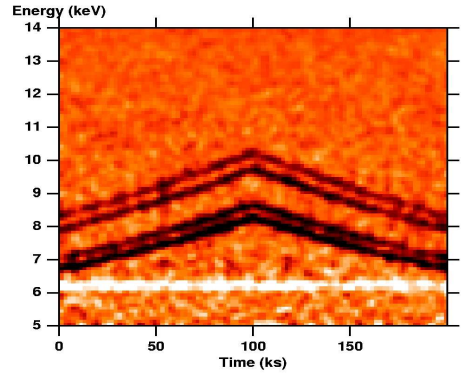}
\caption{\footnotesize The Simbol-X simulation of the time-energy map expected from a source 
with the same spectrum of NGC1365 (Risaliti et al. 2005), i.e. 4 variable absorption lines (from K$\alpha$ and K$\beta$ of 
FeXXV and XXVI; dark lines between 7-10 keV) plus a constant FeK emission line (white line at 6.4 keV). 
The time and energy bins are respectively 2 ks and 100 eV. The total exposure time is 200 ks. The highly 
ionized absorber is assumed to accelerate and decelerate between velocities of 0.02-0.2 c in 100 ks. 
This clearly shows that Simbol-X would be capable of detecting the outflow time evolution 
in bright AGNs, through detailed absorption lines spectroscopy.
}
\label{eta}
\end{figure}

Finally, Simbol-X will be able to detect moderately strong (EW $\lsimeq$ -25 eV) 
absorption lines or edge-like structures at energies up to $\sim$12-15 keV. 
Considering that Fe edges and absorption lines are expected 
at rest-frame energies between E$\sim$7.1-9 keV, Simbol-X will be able to probe 
ionized absorbing gas with velocities up to $\sim$0.5 c. 
This is illustrated in Figure 6 in the case of the bright (F$_{(0.5-10 keV)}$ = 10$^{-11}$ erg 
cm$^{-2}$ s$^{-1}$) quasar PDS456 which, assuming the best-fit model obtained by Reeves et al. (2003), 
could well exhibit edge-like structures from 0.1~c up to 0.5 c.

\begin{figure}[!]
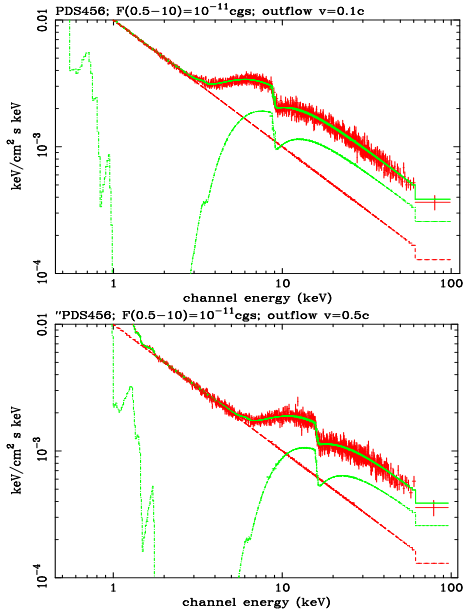

\includegraphics[width=4truecm,height=6truecm,angle=-90]{cappi_fig6a.ps}
\includegraphics[width=4truecm,height=6truecm,angle=-90]{cappi_fig6b.ps}
\caption{\footnotesize Simbol-X simulations for the quasar PDS456, including both MPD (from 0.1 to 30 keV) 
and CZT (from 4 to 100 keV) detectors, for an exposure time of 100 ks. 
The best-fit model consists of a ionized absorber with $\Gamma$=2, 
N$_{\rm W}$=5$\times$10$^{23}$ cm$^{-2}$, log$\xi$=2.6, and a covering factor of 0.6.
We assumed an outflow velocity of ({\it Top}) 0.1 c (similar to what found by Reeves et al. 2003), and 
a more extreme case ({\it Down}) with velocity of 0.5 c.
}
\label{eta}
\end{figure}

\section{Conclusions} 

Recent Chandra and XMM-Newton observations have revealed the presence in bright 
nearby Seyfert galaxies and more distant QSOs of high-velocity, ionized variable absorbers. 
This phenomenon is of outstanding interest because it offers new potential to probe the 
dynamics of the innermost regions of accretion flows, to probe the formation regions of 
outflows and jets, and to help constraining the rate of kinetic energy injected by 
AGNs into the ISM and IGM. 
We have shown that, thanks to its unprecedented sensitivity in the 2-20 keV energy band, 
Simbol-X may lead to remarkable new results in this reasearch field.

\begin{acknowledgements}

We are grateful to M. Dadina, P. Grandi, G. Malaguti and G. Ponti for useful discussions.
M.C. and F.T. acknowledge financial support from ASI under contract ASI/INAF/023/05/0.

\end{acknowledgements}

\bibliographystyle{aa}

\end{document}